\documentclass[preprint,showpacs,preprintnumbers,nofootinbib]{revtex4}
\usepackage{graphicx}
\usepackage{dcolumn}
\usepackage{bm}
\usepackage{amsfonts}

\newcommand{\bea}{\begin{eqnarray}}
\newcommand{\eea}{\end{eqnarray}}
\newcommand{\be}{\begin{equation}}
\newcommand{\ee}{\end{equation}}

\newcommand{\fp}{\phi^{\prime}}

\begin{document}
\preprint{Brown-HET-1398}
\title{Effective Field Theory Approach to String Gas Cosmology}
\author{Thorsten Battefeld}
\email{Battefeld@physics.brown.edu}
 \affiliation{Physics Department, Brown University,
  Providence RI 02912 USA.}
\author{Scott Watson}
\email{watson@het.brown.edu} \affiliation{Physics Department,
Brown University,
  Providence RI 02912 USA.}
\date{\today}

\begin{abstract}
We derive the $4D$ low energy effective field theory for a closed
string gas on a time dependent FRW background. We examine the
solutions and find that although the Brandenberger-Vafa mechanism
at late times no longer leads to radion stabilization, the radion
rolls slowly enough that the scenario is still of interest.  In
particular, we find a simple example of the string inspired dark
matter recently proposed by Gubser and Peebles.
\end{abstract}
\maketitle

\section{Introduction}

String theory continues to be our leading candidate for a quantum
theory of gravity.  However, it continues to be a challenge to
find phenomenological predictions that can be verified by
experiment.  One challenge that stymies the effort towards string
phenomenology is our lack of understanding of the ground state of
string theory.  This has come to be known as the cosmological
moduli problem (see e.g. \cite{dine} and references within). One
approach to resolving this problem is to take the point of view
that cosmological evolution should be responsible for determining
the value of the moduli.  One often finds that, by adopting this
view, the moduli relax to special locations in the stringy
landscape, which are points of enhanced symmetry. A recent example
of such a scenario was presented in \cite{stanford}. There it was
found that the moduli are trapped in orbits around points of
enhanced symmetry due to the production of light string modes.
Stability then sets in due to the Hubble damping resulting from
cosmological evolution. Alternatively, one can also have so-called
racetrack models where the moduli continue to roll and do not
remain fixed, but could give interesting cosmological
consequences \cite{quevedo}.  In this latter approach an important issue is fine
tuning or the cosmic coincidence problem, i.e. why did the moduli
start rolling at such a particular time?

The Brandenberger-Vafa scenario (BV scenario) \cite{bv}, or Brane
Gas Cosmology (BGC)
\cite{loitering,branes,stable,isotropy,perturbations,columbia,columbia2,subodh}
as it has come to be known, is an example of a cosmological model
that incorporates both of these approaches to the moduli problem.
In these scenarios one generally works in the background of ten
dimensional dilaton gravity\footnote{Although this has been
extended to M-theoretical considerations in \cite{columbia2}.},
with sources given by the string winding and momentum modes.  It
is found that the scale of the extra dimensions (radion) is then
stabilized at the self-dual radius, where many of the string modes
become massless and the symmetry is enhanced \cite{stable}\footnote{We note that the analysis of \cite{stable}
was performed with the case of bosonic strings in mind.
Stabilization at the self-dual radius leading to enhanced gauge
symmetry is expected for heterotic strings, but this does not
occur for Type II strings. We thank Steve Gubser for reminding us
of this point.}. A
crucial aspect of these findings was the running of the dilaton to
weak coupling, which was driven by the winding and momentum modes
of the string.  In addition, it has been shown that this model is
stable to both anisotropies \cite{isotropy} and inhomogeneities at
the linear level \cite{perturbations}.

In this paper we would like to extend the BV scenario to better
understand its predictions for late-time cosmology.  In
particular, we would like to see if the stabilization mechanism is
still plausible in the $4D$ effective field theory resulting from
dimensional reduction.  Here, one usually assumes that the dilaton
is fixed, since otherwise this would lead to
unacceptable observational consequences\footnote{However there
have been attempts to keep the dilaton dynamical, see e.g.
\cite{polyakov}}. Given that at late times we are no longer in the
regime of dilaton gravity, perhaps one would naively expect that
such stabilization would no longer work.  That is, in General
Relativity one must generally introduce exotic matter and/or
violate the weak energy condition to stabilize extra dimensions
\cite{Carroll}. The string modes that we will consider here do not
have such properties. To spare the reader
suspense, we do in-fact find that stabilization fails, except in
the special case of one extra dimension\footnote{We mention that the case of $d=1$,
from the perspective of the $5D$ Einstein frame, has already
been considered in \cite{subodh}.}. However, this is not as
disastrous as one might first imagine. In fact, although the
radion is no longer stable in the effective theory, its evolution
is slow enough compared to that of the $4D$ background to be
observationally acceptable.  In addition, we find that this
evolution can lead to interesting phenomenology. As an example, we
find an example of a cold dark matter candidate like that recently
purposed by Gubser and Peebles \cite{gubser}.

In Section II, we will briefly review the radion stabilization
mechanism as presented in \cite{stable}.  We present the stress
energy tensor for the string modes and the corresponding action
for the string modes. In Section III, we consider the evolution in
the Einstein frame.  This is not the correct frame at early times
when one is interested in the geodesics followed by the strings,
but will be important for the late-time cosmology.  In Section IV,
we dimensionally reduce the theory and find a form for the string
modes in the $4D$ effective theory, which is given by their
effective potential. From the effective potential we are able to
discuss the stability of the radion, which is shown to exhibit
slow rolling behavior.  This leads us to the possibility of closed
strings in the extra dimensions behaving as dark matter.

\section{String Gases in the string Frame}
Our starting point is the action
\be
\label{action} S=\frac{1}{2 \kappa_{10}^2}\int d^{D}x
\sqrt{-G} e^{-2 \phi}\Bigl( R+4(\nabla
\phi)^{2}-\frac{1}{12}H^{2} \Bigr)\,,
\ee
which is the low energy effective action for the bosonic degrees of freedom of string theory \cite{polchinski}.
For simplicity, we will ignore flux ($H=0$) in the remainder of this
paper\footnote{See \cite{flux} for a treatment where flux was considered.}.

We wish to consider space-times of the form \be \label{themetric2}
ds^2={G}_{MN}dX^M dX^N={g}_{\mu \nu}(x) dx^{\mu} dx^{\nu} +
{b}^2(x) \gamma_{mn}(y) dy^m dy^n\,, \ee where $x^{\mu}$ are
coordinates in the four dimensional space-time and $y^m$ are the
coordinates of the $d$ extra compact dimensions.  The scale factor
$b(x)$ is the volume modulus of the extra dimensions, which we
will take to be strictly a function of time.  For realistic string
theory compactifications it is necessary that the Ricci curvature
of $\gamma_{m n}(y)$ vanishes, i.e. $R[\gamma_{ij}]=0$ (See e.g.
\cite{witten}). For simplicity, in this paper we will consider the
case of a toroidal background $\gamma_{m n}(y)=\delta_{m n}$. We
will take ${g}_{\mu \nu}$ to be that of a standard, flat FRW
universe and (\ref{themetric2}) can be parameterized as \be
\label{themetric} ds^2=-dt^2+a^2(t) \; d^2\vec{x} + {b}^2(t) \;
d^2\vec{y}\,. \ee

We now want to consider the effect of a gas of string winding and
momentum modes on the cosmological evolution.  We can include
the string modes by adding a matter term to the action
(\ref{action})
\bea \label{matteraction}
S_m&=&\int d^{D}x \sqrt{-G} \rho\,,\\
{T}_{MN}&=&-\frac{2}{\sqrt{{-G}}} \frac{\delta {S}_m}{\delta
{G}^{MN}}\,,
\eea
where $\rho$ is the total energy density of the winding and
momentum modes.
The stress energy tensor of the string gas can be written as
\be
T^{MN}=(T_3^{w})^{MN}+(T_d^{w})^{MN}+(T_3^{m})^{MN}+(T_d^{m})^{MN}\,,
\ee
where these terms represent the separate gases of winding and momentum modes in
both the three large and $d$ small dimensions.
Note that, since we are ignoring interactions between winding and momentum modes and
between strings associated with the separate subspaces, all species are separately conserved.
We should clarify that by {\em interactions} we are referring to the
long-range forces that result, for example, from the gravitational
exchange.  String intersections are of course not possible, since
strings do not generically intersect in the full $10D$ space-time.

Keeping this in mind, we now consider the components of the
{\em total}
stress tensor
\bea \label{stressterms}
-T^0_0 =\rho&=&\frac{d\mu N_d}{\sqrt{G_s}} \;b(t)+\frac{d\mu M_d
}{\sqrt{G_s}}\; b^{-1}(t)
+\frac{3\mu N_3}{\sqrt{G_s}} \;a(t)+\frac{3\mu M_3}{\sqrt{G_s}}\; a^{-1}(t)\,,\nonumber\\
T^i_i=p_i&=&\frac{-\mu N_3}{\sqrt{G_s}} \;a(t)+\frac{\mu M_3}{\sqrt{G_s}}\; a^{-1}(t)\,, \nonumber \\
T^m_m=p_m&=&\frac{-\mu N_d}{\sqrt{G_s}} \;b(t)+\frac{\mu M_d}{\sqrt{G_s}}\; b^{-1}(t)\,,
\eea
where $\sqrt{G_s}=a^3(t) b^d(t)$ is the determinant of the spatial part of the metric.
The equations that follow from the action (\ref{action}) with
the above sources were studied in detail in \cite{stable}.  There it was
shown that if the size of the extra dimensions is taken initially to be on the order of the string
scale, the string modes and dilaton will drive the radion to
the self-dual radius ($b=1$ in string units), while the
other three large spatial dimensions
continue to expand.  This can be understood by considering that the
winding modes become more massive as the extra dimensions expand
and the momentum modes become more massive as they contract.
The self-dual radius is chosen as the dynamically favored scale,
since the mass is minimized there.  Furthermore, many of these
massive string modes actually become massless at the self-dual
radius, which is a location of enhanced gauge symmetry.
This is an example of the Higgs mechanism in string
theory, where the role of the Higgs is played by the radion
or scale of the extra dimensions \cite{polchinski}.

We now wish to see if this argument survives in the lower
dimensional effective field theory.  This is crucial for the BV
scenario, since once the three dimensions have grown large enough, the small
dimensions can be integrated out and the theory should be
described by a $4D$ low energy effective field theory. But before
dimensionally reducing, we would like to discuss the evolution in
the Einstein frame.

\section{String Gases in the Einstein Frame}
We can move to the Einstein frame by performing a conformal rescaling
of the metric and a field redefinition of the dilaton,
\be \label{conformal}
\tilde{G}_{MN}=e^{-\frac{4 \phi}{(2+d)}}G_{MN} \,\,\,\,\,,\,\,\,\,\,
\tilde{\phi}=\frac{4 M_D}{\sqrt{2+d}} \phi\,,
\ee
where quantities with a {\em tilde} refer to the Einstein frame
metric. In terms of the new fields the action (\ref{action}) becomes
\be \label{eframeaction1}
S_{E}= \int d^{D}x \sqrt{-\tilde{G}} \Biggl( \frac{1}{2} M_D^2\tilde{R} - \frac{1}{2}
\tilde{G}^{MN}\tilde{\nabla}_M \tilde{\phi}\tilde{\nabla}_N \tilde{\phi}
\Biggr)+\tilde{S}_m\,,
\ee
where we take $M^{-2}_D=8 \pi G_D$ as the $D$ dimensional Planck
mass. The matter action is now given by
\bea \label{cmatteraction}
\tilde{S}_m&=&\int d^{D}x \sqrt{-\tilde{G}} \tilde{\rho}\,,
\eea
where the energy density $\tilde{\rho}$ is not simply the transformed density $\rho$ from the
string frame, since one has to take the transformation of the determinant of the metric into account, that is
\begin{eqnarray}
\sqrt{-G} &=& \sqrt{-\tilde{G}} e^{\frac{4+d}{2\sqrt{2+d}} \bigl( \frac{\tilde{\phi}}{M_D}\bigr)}\,.
\end{eqnarray}
Our energy and pressure terms are then given in terms of the
transformed stress energy tensor
\bea \label{estressterms}
-\tilde{T}^0_0 =\tilde{\rho}&=&\frac{d\mu N_d}{\sqrt{\tilde{G_s}}} \;\tilde{b}
e^{\frac{1}{\sqrt{2+d}} \bigl( \frac{\tilde{\phi}}{ M_D}\bigr)}+\frac{d\mu
M_d
}{\sqrt{\tilde{G_s}}}\; \tilde{b}^{-1} +\frac{3\mu N_3}{\sqrt{\tilde{G_s}}} \;\tilde{a}
e^{\frac{1}{\sqrt{2+d}}
\bigl( \frac{\tilde{\phi}}{ M_D}\bigr)} +\frac{3\mu
M_3}{\sqrt{\tilde{G_s}}}\; \tilde{a}^{-1} \,, \nonumber\\
\tilde{T}^i_i=\tilde{p}_i&=&-\frac{\mu N_3}{\sqrt{\tilde{G_s}}} \;\tilde{a}
e^{\frac{1}{\sqrt{2+d}}
\bigl( \frac{\tilde{\phi}}{ M_D}\bigr)} +\frac{\mu
M_3}{\sqrt{\tilde{G_s}}}\; \tilde{a}^{-1}\,,  \nonumber \\
\tilde{T}^m_m=\tilde{p}_m&=& -\frac{\mu N_d}{\sqrt{\tilde{G_s}}} \;\tilde{b}
e^{\frac{1}{\sqrt{2+d}} \bigl( \frac{\tilde{\phi}}{ M_D}\bigr)}+\frac{\mu
M_d
}{\sqrt{\tilde{G_s}}}\; \tilde{b}^{-1}\,.
\eea
Finally, the equations of motion are given by
\begin{eqnarray}
\tilde{R}_0^0-\frac{1}{2}{\delta}_0^0\tilde{R}&=&\frac{1}{M_D^2}\left(\tilde{\rho}+\frac{1}{2}\dot{\tilde{\phi}}^2\right)\,,\label{EOM10D1}\\
\tilde{R}_\mu ^\mu -\frac{1}{2}{\delta}_\mu ^\mu\tilde{R}&=&-\frac{1}{M_D^2}\left(\tilde{p}_i+\frac{1}{2}\dot{\tilde{\phi}}^2\right)\,,
\\
\tilde{R}_m^m-\frac{1}{2}{\delta}_m^m\tilde{R}&=&-\frac{1}{M_D^2}\left(\tilde{p}_m+\frac{1}{2}\dot{\tilde{\phi}}^2\right)\,,\\
\tilde{\Box} \; \tilde{\phi}&=&
-\frac{d\tilde{\rho}}{d\tilde{\phi}}\label{EOM10D4}\,,
\end{eqnarray} where it is understood that repeated indices are
not summed over.
We will not be interested in exact solutions to the above
equations at this point.  However, we would like to pause and make some important
comments regarding their interpretation.

\subsection{Switching Frames and Stability}
It is often argued that by
moving to the Einstein frame, dilaton gravity is simply general relativity
with the addition of scalar matter.  That is indeed true for the vacuum case, however
when string sources are present the dilaton couples differently to winding and momentum modes.
This is manifested by our stress energy terms in (\ref{estressterms}).
For example, if we substitute the energy density in (\ref{estressterms}) back into
the matter action (\ref{cmatteraction}), we see that momentum modes are conformally
invariant (as expected since they behave like radiation, which has
a traceless stress energy tensor), but the winding modes are affected by
the conformal transformation.  This complicated coupling to the dilaton is just an example
of why the Einstein frame is the unnatural frame for interpreting the
physics at early times\footnote{This has been discussed much in the literature,
see for example \cite{deflation}.}. However, here we are interested in making contact with
late-time cosmology, when the Einstein frame becomes relevant.

One can understand the difference in the frames quantitatively by
examining the Einstein frame line element in terms of the string
frame scale factors
\be
d\tilde{s}_E^2=-e^{-\frac{4\phi}{(2+d)}}dt^2+
e^{-\frac{4\phi}{(2+d)}}a^2(t)d^2\vec{x}+e^{-\frac{4\phi}{(2+d)}}b^2(t)d^2\vec{y}\,.
\label{argumentstable} \ee Recalling that the evolution of the
dilaton was crucial for stabilizing the scale factor $b(t)$ in the
string frame, we can immediately see the difficulty in stabilizing
the Einstein frame scale factor
$\tilde{b}(t)=e^{-\frac{2\phi(t)}{(2+d)}}b(t)$.  That is, we see
that an evolving dilaton is in contradiction with the
stabilization of $\tilde{b}(t)$. One way to resolve this is to fix
the dilaton at late-times, such that the scale factors are fixed
at the self-dual radius $b(t)=\tilde{b}(t)=1$ in both frames. This
argument is intuitively correct, since once the dilaton is fixed
the string and Einstein frames are equivalent.  However, we will
see in the next section that fixing the dilaton will not be enough
to maintain stabilization for the case of more than one extra
dimension.

The reader may argue that one needs a specific mechanism to
provide a potential for the dilaton.  It is generally believed that this
mechanism should come from a better understanding of the nonperturbative theory
and is perhaps related to the breaking of supersymmetry.  We do not wish to address
this issue here, except to comment that our conclusions seem to be
robust in this regard, as long as
we remain in the weak coupling regime.
For example, one could imagine trying to incorporate the scenario
we will describe below in a Type
IIB construction with fluxes, where all moduli
(including the dilaton) have been fixed accept for the radion
\cite{stanfordinflation}.  Another possibility is to follow the
point of view of \cite{polyakov} and keep the dilaton dynamic.

In what follows, we will consider the stabilization in the ten
dimensional string frame to be accurate at high energy/small
distances.  Therefore, we accept the results of \cite{stable} as
being accurate in this early regime.  Then, once we reach the stage of
late-time cosmology we assume that the issue of SUSY symmetry breaking
and the evolution or fixed dilaton are given, {\em a priori}.
Following this cosmological transition we expect the physics to be
described by the $4D$ effective field theory.  We will now discuss
what influence the closed string modes have in this regime.

\section{String Gases in Four Dimensional Einstein Gravity}
We now want to consider the four dimensional effective
field theory resulting from the compactification of the $d$
internal dimensions \cite{Carroll}.  We can dimensionally reduce
the action (\ref{eframeaction1}) by decomposing the determinant of
the metric (\ref{themetric}) and the Ricci scalar as \bea
\sqrt{-\tilde{G}}&=&\tilde{b}^d\sqrt{-\tilde{g}}\,,\\
R[\tilde{G}_{MN}]&=&R[\tilde{g}_{\mu
\nu}]-2d\tilde{b}^{-1}\tilde{g}^{\mu
\nu}\tilde{\nabla}_{\mu}\tilde{\nabla}_{\nu}\tilde{b}
-d(d-1)\tilde{b}^{-2}\tilde{g}^{\mu
\nu}\tilde{\nabla}_{\mu}\tilde{b}\tilde{\nabla}_{\nu}\tilde{b}\,,
\eea where we have again used $R[\gamma_{ij}]=0$. The
compactification scale is usually taken to be around the Planck
scale. This means that one can integrate out the dependence of the
fields on the coordinates of the extra dimensions. We obtain the
low energy effective action for the four dimensional theory \bea
\label{effectiveaction} S_{eff}&=&\int d^{4}x \sqrt{-\tilde{g}} \;
\Biggl[  \frac{1}{2}M_p^2 \Bigl(\tilde{b}^d \tilde{R}[\tilde{g}
_{\mu \nu}] -2d\tilde{b}^{d-1}\tilde{g}^{\mu \nu}
\tilde{\nabla}_{\mu}\tilde{\nabla}_{\nu}\tilde{b}
-d(d-1)\tilde{b}^{d-2}\tilde{g}^{\mu \nu}
\tilde{\nabla}_{\mu}\tilde{b} \tilde{\nabla}_{\nu}\tilde{b}\Bigr)
\nonumber\\ &&-V_d\; \tilde{b}^d \tilde{g}^{\mu \nu}
\tilde{\nabla}_{\mu} \tilde{\phi}\tilde{\nabla}_{\nu} \tilde{\phi}
 + V_d \; \tilde{b}^d \; \tilde{\rho} \Biggr]\,,
\eea
where we have defined $V_d=\int d^dy \sqrt{\gamma}$
as the spatial volume of the extra dimensions for unit scaling ($\tilde{b}=1$) and the
four dimensional Planck mass is given by $M_p^2=V_d M_{D}^2$.

We now perform another conformal rescaling and field redefinition to put the action
(\ref{effectiveaction}) in canonical form,
\be \label{f1}
\psi=\sqrt{\frac{d(2+d)}{2}}M_p \ln{(\tilde{b})}, \;\;\;\;\;\;\;\;\;
\bar{g}_{\mu \nu}=e^{\sqrt{\frac{2d}{2+d}} \; \bigl(\frac{\psi}{M_p}\bigr) } \; \tilde{g}_{\mu
\nu},\;\;\;\;\;\;\;\;\; {\varphi}=V_p^{1/2}\tilde{\phi}\,,
\ee
where we have introduced the four dimensional dilaton $\varphi$.
We parametrize our metric by
\be \label{bgcmetric}
ds^2=\bar{g}_{\mu \nu} dx^{\mu} dx^{\nu}=-dt^2+e^{2 \lambda(t)}d\vec{x}^{\: 2}, \;\;\;\;\;\;\;
\lambda(t)=\ln{\bar{a}(t)}\,.
\ee
Then the four dimensional effective action with canonically normalized
fields is given by
\bea \label{theaction}
S_{eff}&=& \int d^4x \sqrt{-\bar{g}} \Biggl( \frac{1}{2} M_p^2 R[\bar{g}_{\mu \nu}]-\frac{1}{2}
\bar{g}^{\mu \nu} \bar{\nabla}_\mu \psi \bar{\nabla}_\nu \psi-\frac{1}{2}
\bar{g}^{\mu \nu}\bar{\nabla}_{\mu} {\varphi} \bar{\nabla}_{\nu}
{\varphi} + V(\lambda,\varphi,\psi) \Biggr),
\eea
where the effective potential is obtained from dimensionally reducing the matter action
(\ref{cmatteraction}), applying the
transformations (\ref{f1}) and plugging in the metric (\ref{bgcmetric}).  One finds
\bea
V(\lambda,\varphi,\psi)&=&\mu V_d e^{\frac{1}{\sqrt{2+d}}\frac{\varphi}{M_p}}
\left( 3N_3 e^{-2\lambda}e^{-\sqrt{\frac{2d}{2+d}}\bigl( \frac{\psi}{M_p}\bigr)}
+dN_d e^{-3\lambda}e^{(1-\frac{d}{2})\sqrt{\frac{2}{d(2+d)}}\frac{\psi}{M_p}}
\nonumber\right) \\
&&+\mu V_d\left(3M_3 e^{-4\lambda}+dM_d e^{-3\lambda}e^{-(1+\frac{d}{2})\sqrt{\frac{2}{d(2+d)}}\bigl(
\frac{\psi}{M_p}\bigr)}\right)\,.\label{potential}
\eea
With this potential, the equations of motion are
\begin{eqnarray}
3 \dot{\lambda}^2&=&\frac{1}{2}\dot{\varphi}^2 +\frac{1}{2}\dot{\psi}^2+V(\lambda,\varphi,\psi)\,,\label{EOM1}\\
2\ddot{\lambda}+3\dot{\lambda}^2&=&-\frac{1}{2}\dot{\varphi}^2-\frac{1}{2}\dot{\psi}^2
+\frac{1}{3\sqrt{\bar{g_s}}}\frac{\partial (\sqrt{\bar{g_s}}\; V)}{\partial \lambda}\,,\label{eom13a}\\
\ddot{\varphi}+3\dot{\lambda}\dot{\varphi}&=&-\frac{\partial V}{\partial \varphi}\,,\\
\ddot{\psi}+3\dot{\lambda}\dot{\psi}&=&-\frac{\partial V}{\partial \psi}\,, \label{EOM3}
\end{eqnarray}
where we set $M_p\equiv 1$ and $\bar{g}_s$ is the determinant of the spatial part
of the metric (\ref{bgcmetric}).  For example, if the dilaton and radion are slowly
rolling and we take $N_3=N_d=M_d=0$ we are left with the equations
for a radiation dominated universe,
\begin{eqnarray}
3 \dot{\lambda}^2&\sim &e^{-4\lambda} \sim \bar{a}^{-4}(t)\,,\\
2\ddot{\lambda}+3\dot{\lambda}^2&\sim & -\frac{1}{3}\bar{a}^{-4}(t)\,,
\end{eqnarray}
as we expect from a gas of massless bosonic strings in the
uncompactified directions.

\subsection{Late time solutions with $d=1$ and dark matter \label{d=1}}
Let us first consider the case of one extra dimension ($d=1$) and
focus on the winding and momentum modes in the extra dimensions
only, i.e. $N_3=M_3=0$.  We are justified in doing this, since at
late times the winding modes have all annihilated in the large
dimensions \cite{isotropy} and as we have seen the momentum modes
behave as a gas of radiation, which is subdominant in the matter
dominated epoch.

To understand the late time evolution, we must decide how to evolve
the dilaton.  Here we choose to give the dilaton
a VEV, which we argued in Section IIIA should be determined by fixing the minimum of the
potential to be at $\psi=0$, i.e. the self-dual radius. Thus, we fix
$\phi_0=0$
and at the self-dual radius one has $N_d \approx M_d$.
The potential is given by

\be \label{d1V}
V(\psi,\lambda)=\frac{\mu V_d N_d e^{\frac{\sqrt{6} \psi}{6
M_p}}}{e^{3\lambda}}+\frac{\mu V_d M_d e^{-\frac{\sqrt{6} \psi}{2
M_p}}}{e^{3\lambda}},
\ee
so that
\be
V(\psi=0,\lambda)\sim \frac{1}{e^{3\lambda}}, \ee which one
recognizes as the energy density for matter. We can take it to be
in the dark sector, because of its stringy origin. This leads to a
simple example of the string inspired dark matter discussed
recently by Gubser and Peebles in \cite{gubser}.  In that paper it
was shown that by considering long-range scalar and gauge
interactions, one is led to slightly modified $\Lambda$CDM models
that may still be within observational bounds.  We refer readers
to that paper for the details, but here we wish to demonstrate
that our naive string gas model reproduces one such dark matter
candidate. The crux of the modification to the $\Lambda$CDM model
is a {\em fifth force} that is provided by the long-range
interactions of the string modes. Recall that winding strings do
not intersect in greater than three dimensions, that is why we
were led to three dimensions decompactifing and the other $d$
remaining compact as motivated by the BV scenario. However, the
strings do interact through their effects on the gravitational
background and this leads to a modification of the local
gravitational force.  For two particles with masses $m_i$ and
$m_j$ ($i=1,2$) the force is given by \cite{gubser},
\be
F_{ij}=\beta_{ij}\frac{Gm_i m_j}{r^2}\,, \;\;\;\;
\beta_{ij}=1+2 M_p^2 \frac{Q_i Q_j}{m_i m_j}\,,
\ee
where the $m_i$ are obtained from the potential (\ref{d1V}) by
\be
V(\psi,\lambda)=\sum_{q}n_i m_i\,,
\ee
so that we identify
\be
n_1=\frac{M_d}{e^{3\lambda}}\,, \;\;\;
n_2=\frac{N_d}{e^{3\lambda}}
\ee
as the number densities and
\be
m_1=\mu V_p e^{-\frac{\sqrt{6}}{2}\frac{\psi}{M_p}}, \;\;
m_2=\mu V_p e^{\frac{\sqrt{6}}{6}\frac{\psi}{M_p}}
\ee
as the masses. Finally, the {\em charges} $Q_i$ are given by
\be
Q_i=\frac{d m_i}{d\psi}\,.
\ee
We also need to introduce the mass fraction
$$f_i=\frac{n_i m_i}{\sum_k n_k m_k},$$ which for our model is
$f_1=f_2=\frac{1}{2}$.

The statement that $\psi=0$ is a stable minimum becomes
that of charge neutrality for the charges $Q_i$, which reads
\be
\sum_i n_i Q_i=0\,,
\ee
with $3M_d=N_d$.

If we calculate
$\beta_{ij}$ for this model we find
\[ \beta_{ij}= 4 \left[
         \begin{array}{rr}
              1 & 0 \\
              0 &  \frac{1}{3}
          \end{array} \right]. \]
First, we note that the off-diagonal entries are zero, so that
winding and momentum modes are not interacting.  This was
our starting assumption, so this offers a nice consistency check.

As discussed in \cite{gubser}, the growth of instability is given in terms of the mass
density contrast $\delta_i=\delta \rho_i / \rho$ where $\rho$ is the total energy density and
\be
\ddot{\delta}_i+2\dot{\lambda} \dot{\delta}_i=\frac{\rho}{2
M_p^2}\sum_j \beta_{ij} f_j \delta_j,
\ee where the second term is a damping factor which expresses the
decay of peculiar velocity due to cosmological expansion.  The
problem of finding the modes of instability is made more tractable
by introducing
\be
\Delta=\sum_i c_i \sqrt{f_i} \delta_i,
\ee where the $c_i$ are constants.  If one chooses these constants
appropriately and introduces the matrix
\be \label{eqnsc}
\Xi_{ij}=\sqrt{f_i} \beta_{ij} \sqrt{f_{j}},
\ee
then $\Delta$ satisfies
\be
\ddot{\Delta}_a+2 \dot{\lambda} \dot{\Delta}_a -\frac{\rho}{2
M_p}\xi_a \Delta_a=0,
\ee
where the $\xi_a$ are the eigenvalues of $\Xi$.  To find the eigenvalues we must multiply
(\ref{eqnsc}) by an extra factor of $\frac{1}{2}$, so that our
normalization of $\psi$ agrees with that of \cite{gubser}.  We
then find the eigenvalues $\xi_1=1$ and $\xi_2=\frac{1}{3}$.  It
then follows that
\be
\Delta_1 \sim t^{2/3}, \;\;\;\;\;\; \Delta_2 \sim t^{1/3},
\ee are the fastest growing modes.
The first is an adiabatic mode, which corresponds to the motion of the string modes moving
together with the expansion in the matter dominated epoch, while the
second mode is subdominant.  Since models of this type were discussed in detail in
\cite{gubser}, we will simply conclude this section by commenting
that our naive string gas model appears to lead to the possibility
of a cold dark matter candidate for $d=1$.  However, one finds
that in the case of more than one extra dimension no local minimum exists for $V(\psi)$, so
that the question of stability becomes more of a sensitive issue, which we now discuss.

\begin{figure}[tb]
  \includegraphics[width=\columnwidth]{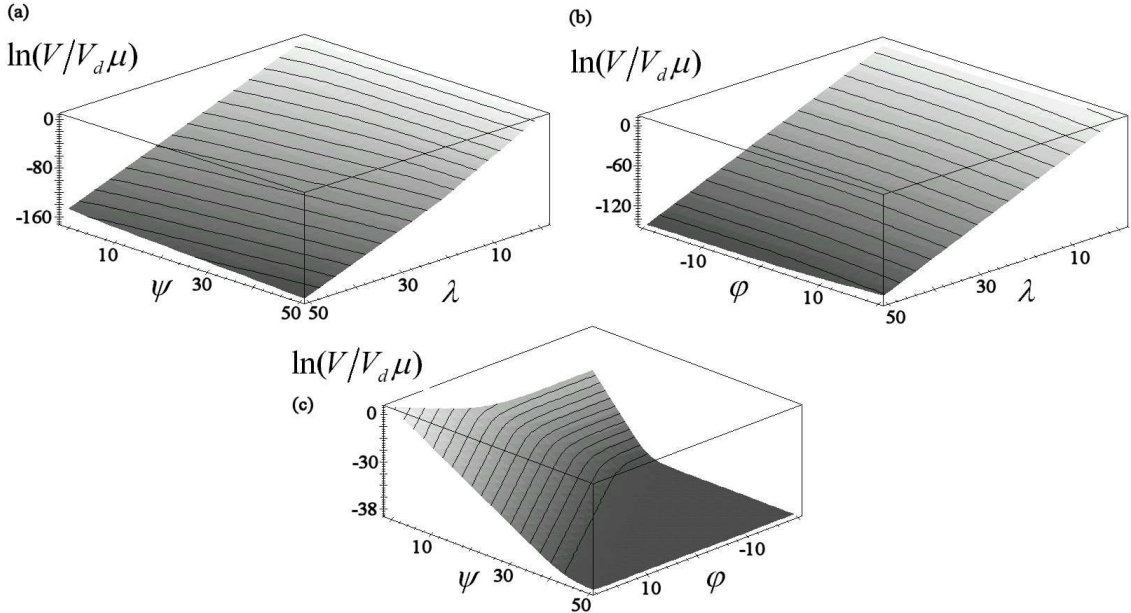}
   \caption{\label{pic:potential} The potential $\ln(V(\lambda,\varphi,\psi)/(V_d\mu))$ from (\ref{potential}) is plotted
   with $d=6$, $N_d=M_d=M_3=1$, $N_3=0$, $M_p\equiv 1$ and (a) the dilaton held constant at $\varphi=-3/M_p$, (b)
   the extra dimensions held constant at $\psi=0$ and (c) the large dimensions held constant
   at $\lambda=10$.}
\end{figure}

\subsection{Late time solutions for $d>1$ \label{d>1}}
We have presented the $d=1$ case as an explicit illustration of the possibility
of obtaining CDM models from the string gas approach.
In the case $d>1$, one can see that the potential (\ref{d1V}) has no local minimum.
This means that once
the effective field theory becomes relevant, one expects
the radion to begin
rolling and the extra dimensions to expand.
This seems to be a negative finding for the stabilization of extra
dimensions at late times, within this naive extension of the BV scenario.
Moreover, the CDM model relies crucially on the vanishing
of the pressure around the minimum $\psi=0$.  Since
the minimum vanishes for $d>1$, one must worry that if the pressure
becomes appreciable we will not be in a matter dominated epoch.

\begin{figure}[tb]
  \includegraphics[width=\columnwidth]{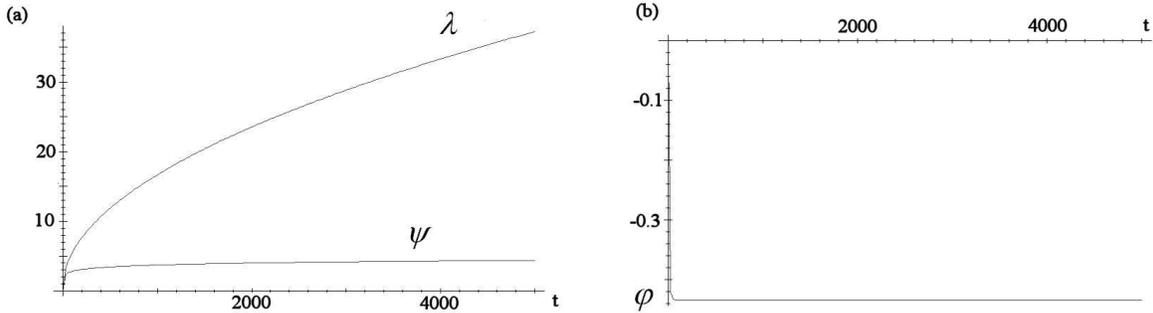}
   \caption{\label{pic:solution} The solution to equations (\ref{EOM1})-(\ref{EOM3})
   for the potential $V(\lambda,\varphi,\psi)$ from (\ref{potential}) as plotted in Fig. \ref{pic:potential}
   with the same settings and the initial conditions $\varphi=\dot{\varphi}=\psi=\dot{\psi}=\lambda=0$:
   (a) evolution of $\lambda(t)$ (large dimensions, upper curve) and $\psi(t)$ (extra dimensions, lower curve),
   (b) evolution of the dilaton $\varphi(t)$.}
\end{figure}

To spite these seemingly negative results, a resolution is immediately
apparent.  If one examines the potential (\ref{potential}), one finds
that radion rolls slowly down its
potential compared to the expansion.  In fact, one finds
\be
\frac{\partial (\ln V)}{\partial \psi}<1, \;\;\;\;\;\;\;\; \frac{\partial^2 V}{\partial
\psi^2}<1\,,
\ee
quite generally.
As a specific example, let us consider the case discussed in \cite{stable}. That
is,
$d=6$ and the winding modes have all annihilated in the large
dimensions, so $N_3=0$.  The resulting potential is plotted in Fig.~\ref{pic:potential}(a-c).
We see that the potential is steepest in the $\lambda$ direction as can be seen in Fig.~\ref{pic:potential}
or computed directly using (\ref{potential}).
Thus, the three dimensions
will grow much faster than the radion.
At the same time the dilaton will roll
to more and more negative values, which is a result of the string modes driving the model to
weak coupling.  The dilaton will grow faster in the early phase but slowing down after $\psi$ has grown to an
appreciable size. This
can be seen in Fig.~\ref{pic:potential}(b,c) by the shallowness in the direction of the dilaton after $\psi$ has grown.
The same generic behavior is found for the case of $d=2,3,4,$ or $5$, since the potential has a similar shape.
Thus, once the
transition takes place the radion can roll slowly enough that the
pressure remains negligible and a dark matter candidate arises as in the $d=1$ case.
In Fig.~\ref{pic:solution} we present a sample numeric solution
to (\ref{EOM1})-(\ref{EOM3}), which exemplifies this slow roll behavior.
We note that this behavior is generic for initial conditions that
respect the condition $g_{\text{string}} \sim e^{2 \phi}\ll 1$.

Although this behavior is promising, a more detailed
investigation is needed.  In particular, one needs to better
understand the dark matter in the case of appreciable pressure.
That is, if the pressure becomes
significant the analysis reviewed in Section IV-A needs to be revised.
One might also entertain the possibility of obtaining dark energy
from such a model, however this seems to fail since the pressure
is positive (leading to contraction) as can be seen from (\ref{eom13a}).
Regardless of what properties
this new matter has, it seems worthy of future study.

\section{Conclusion}
In this article we have
shown that stabilization of the radion in the effective theory at
late-times is problematic.  This was not the case with one extra dimension, where after
passing to the effective theory the potential retains a minimum at the self-dual radius.
In that case, we were able to construct a simple realization of scalar dark matter as discussed in
\cite{gubser}. However, in the case of more than one
extra dimension the potential does not possess a local minimum.
This led to extra dimensions that are slowly growing, but that remain small compared to the large
dimensions.  This case is interesting in
many respects, e.g. it could lead to {\em large} extra dimensions
without the need to invoke brane world scenarios.
We also note that these results remain valid for both a fixed and evolving dilaton, where
in the later case the string modes drive the dilaton towards the region of weak coupling.
We find the interpretation of winding and momentum modes as CDM
much more involved, but possible.  We leave such considerations
for future work.

\begin{acknowledgments}
We wish to thank Steve Gubser, David Lowe, and
Subodh Patil for useful comments and discussions.  We would also like to thank
Robert Brandenberger and Sera Cremonini for comments on the manuscript.
SW wishes to acknowledge NASA GSRP.  This work was also supported in part by
the U.S. Department of Energy under Contract DE-FG02-91ER40688,
TASK A.
\end{acknowledgments}


\end{document}